\documentstyle[prl,aps,twocolumn,epsf]{revtex}
\begin{document}

\title{\hspace{9cm} {\it to be published in} Physics Letters A\\
\vskip 1.5cm Universal Temperature Behavior of Remanent Magnetization
Observed in Low-$T_c$ and High-$T_c$ Josephson Junction Arrays}

\author{S. Sergeenkov$^a$, W.A.C. Passos$^b$, P.N. Lisboa-Filho$^b$,
and W.A. Ortiz$^b$}
\address{
$^a$Joint Institute for Nuclear Research, Bogoliubov Laboratory of
Theoretical Physics, 141980 Dubna, Moscow region, Russia}
\address {$^b$Grupo de Supercondutividade e Magnetismo,
Centro Multidisciplinar para o Desenvolvimento de Materiais
Ceramicos, Departamento de F\'{i}sica, Universidade Federal de
S\~{a}o Carlos, Caixa Postal 676 - 13565-905 S\~{a}o Carlos, SP,
Brazil }
\address{\em (\today)}
\draft \maketitle
\begin{abstract}
A comparative study of the magnetic remanence exhibited by
tridimensional Josephson junction arrays in response to an excitation
with an AC magnetic field is presented. The observed temperature
behavior of the remanence curves for disordered arrays fabricated
from three different materials ($Nb$, $YBa_2Cu_3O_{7-\delta}$ and
$La_{1.85}Sr_{0.15}CuO_{4-\delta}$) is found to follow the same
universal law (based on the explicit temperature expressions for the
activation energy and the inductance-dominated contribution to the
magnetization of the array within the framework of the phase-slip
model) regardless of the origin of the superconducting electrodes of
the junctions which form the array.
\end{abstract}

 \narrowtext

 As it has been recently found\cite{1,2,3}, tridimensional disordered
Josephson junction arrays (JJAs) fabricated from either low-$T_c$
(LTS) or high-$T_c$ (HTS) superconductors may, upon excitation by a
magnetic field, exhibit a temperature-dependent magnetic remanence, $
M_R(T)$. Typically\cite{4}, the magnetized state occurs in a rather
narrow window of temperatures, the extent of which depends on the
critical current, $I_c(T)$, of the junctions. Besides, there is a
threshold value for the magnetic field in order to drive the JJA to
the state where flux is retained after suppression of the
field\cite{4}.

In this Letter we present a comparative study of three different
samples with a rather spectacular remanent behavior and suggest a
possible interpretation of the observed temperature dependence of the
remanent magnetization of both LTS and HTS tridimensional disordered
JJAs. Our analysis shows that all the experimental data can be rather
well fitted using the explicit temperature expressions for the
activation energy and the inductance-dominated contribution to the
magnetization of the array within the so-called phase-slip
model\cite{5,6,7,8,9}.

Three samples were prepared from selected material, respectively of
$Nb$, $YBa_2Cu_3O_{7-\delta}$ (YBCO) and
$La_{1.85}Sr_{0.15}CuO_{4-\delta}$ (LSCO). All three exhibit the
predicted remanence and other characteristic features of Josephson
arrays. Fabrication routes as well as the experimental routines
employed for the magnetic measurements are described
elsewhere\cite{1,2,3}. In short, the corresponding (e.g., niobium)
powder was separated according to grain size (using a set of special
sieves, with mesh gauges ranging from $38$ to $44 \mu m$), then
uniaxially pressed in a mold to form a cylindrical pellet of $2.5 mm$
radius by $2.0 mm$ height. This pellet is a tridimensional disordered
JJA in which the junctions are weakly-coupled grains, i.e.,
weak-links formed by a sandwich between ($Nb$) grains and a
($Nb$-oxide) layer originally present on the grain surface.

The measurements were made using a Quantum Design MPMS-5T SQUID
magnetometer featuring the regular DC extraction magnetometer and an
AC-susceptibility module. The remanence was obtained measuring the
sample magnetization after application and removal of a train of
sinusoidal pulses. Using the field scan routine we measured the
remanent magnetization as a function of the excitation field. For an
ordinary superconductor of any kind, from a single crystal to a
totally disordered granular sample, the only possibility of a
remanence after the application of the AC field would be a residual
magnetization due to flux eventually pinned inside the specimen. This
contribution, however, is expected to be small and practically
independent of the excitation field. We have verified the above
characteristics measuring $M_{R}(h, T)$ for a variety of samples. In
particular, the powder used to fabricate our arrays have the typical
response of ordinary superconductors, so that the effects described
below are entirely due to the formation of the 3D-JJA. The
experimental results for all three samples (along with the model
fits, see below) are summarized in Fig.1 which suggests that the
observed behavior seems to follow a universal temperature pattern,
irrespective of the type of superconductor of which the array is
made. Let us turn to a possible interpretation of the obtained
results.

Since the observed remanent magnetization (RM) in our samples (JJAs)
appears only below the so-called phase-locking temperature $T_J$
(which marks the establishment of phase coherence between the
adjacent grains in the array and always lies below a single grain
superconducting temperature $T_C$), it is quite reasonable to assume
that origin of RM is related to thermal fluctuations of the phases of
the superconducting order parameters across an array of Josephson
junctions (the so-called phase-slip mechanism\cite{5,6,7,8,9}). In
the present approach we consider the sample as a single plaquette
with four Josephson junctions (JJs), each of which is treated via an
effective single junction approximation. Within this approximation,
the phase-slip scenario yields then
\begin{equation}
\Delta M_R(T)\equiv M(T) - M_R = M_0(T){I_0^{-2}[\gamma (T)/2]} - M_R
\end{equation}
for the observed remanent magnetization. Here, $M_0(T)$ is an
inductance-induced contribution to the magnetization of the array
(see below), $\gamma (T) = U(T)/k_BT$ is the normalized barrier
height for thermal phase slippage, $I_0(x)$ is the modified Bessel
function, and $M_R = M(T_J)$ is a residual temperature-independent
contribution (notice that, according to Eq.(1), $\Delta M_R(T_J) =
0$).

For temperatures below $T_J$ (where the main events take place, see
Fig.1), the Bessel function can be approximated as $I_0(x)\simeq e^x
/\sqrt{2\pi x}$ leading to a simplified version of Eq.(1):
\begin{equation}
M(T) = 2\pi M_0(T)[U(T)/k_BT]\exp[-U(T)/k_BT]
\end{equation}
Figure 1 shows the temperature dependence (in reduced units, $\tau
=T/T_J$) of the normalized remanent magnetization $m_r(T) = \Delta
M_R(T)/\Delta M_R(T_p)$, where $T_p$ is the peak temperature and
$\Delta M_R(T)$ is defined via Eqs. (1) and (2). The data for YBCO-
and Nb-based JJAs are found to be well fitted with the following
explicit expression for the array magnetization:
\begin{equation}
 M(t) = A(1-t^4)^{5/2}\exp[-\alpha (1-t^4)]
\end{equation}
where $t = T/T_C$. The best fits through all the data points (shown
in Fig.1 by solid and dotted lines for YBCO- and Nb-based JJAs,
respectively) using Eq.(3) and the known critical parameters
\begin{equation}
YBCO: \quad T_C = 90 K , \quad T_J = 82 K, \quad T_P = 0.88T_J;
\end{equation}
\begin{equation}
Nb: \quad T_C = 9.1 K , \quad T_J = 8.2 K, \quad T_P = 0.92T_J;
\end{equation}
yield the following estimates of the model parameters: $\alpha
_{YBCO} = 7$ and $\alpha _{Nb} = 9$.

At the same time, the data for the LSCO sample (which appears to have
links weaker than the other two samples) are found to be better
fitted using the following expression:
\begin{equation}
 M(t) = B(1-t^2)^{5/2}\exp[-\beta (1-t^2)^{3/2}],
\end{equation}
with
\begin{equation}
LSCO: \quad T_C = 36.5 K , \quad T_J = 19.87 K, \quad T_P = 0.7 T_J;
\end{equation}
yielding $\beta _{LSCO} = 2$.

To understand the observed behavior of the remanent magnetization, we
need to specify the temperature dependencies of the activation energy
$U(T)$ and the inductance-dominated contribution $M_0(T)$ to the
magnetization of the array. Starting with the YBCO- and Nb-based
arrays, it is reasonable to assume that\cite{10,11} $U(T) = \Phi _
0I_C(T)/2\pi$ and $M_0(T) = LI_C(T)/\mu _0S$, where $I_C(T)$ is an
average value of the critical current, $L$ is an average inductance
of the Josephson network, $S$ is an effective (in general,
temperature-dependent, see below) projected area of the contact,
$\Phi _0$ is the flux quantum, and $\mu _0$ is the vacuum
permeability. In turn, the temperature dependence of the critical
current is dictated by the corresponding dependence of the London
penetration depth, namely\cite{8}:
\begin{equation}
I_C(T) = I_C(0)[\lambda _L(0)/\lambda _L(T)]^2
\end{equation}
where
\begin{equation}
\lambda _L (T) = \lambda _L (0)[1-(T/T_C)^4]^{-1/2}
\end{equation}
Finally, to arrive at the fitting expression given by Eq.(3), we have
to assume that the projected area $S$ is also temperature dependent
(which is not unusual), viz. $S(T) = \pi d(T)l$ with $d(T)$ and $l$
being the thickness and the length of a SIS-type sandwich,
respectively ($d(T) = 2\lambda _L(T) + \xi$, where $\lambda _L(T)$ is
the London penetration depth and $\xi$ is the thickness of an
insulating layer; in ceramics $l$ plays the role of an average grain
size $r_g$; typically, $l \gg \lambda _L(T) \gg \xi$).

Turning now to the LSCO-based array, we notice that in this case the
temperature dependence of the activation energy $U(T)$ is governed by
the well-known weak-links mediated expression\cite{8} $U(T) =
U(0)[1-(T/T_C)^2]^{3/2}$, while the critical current reads $I_C(T) =
I_C(0)[1-(T/T_C)^2]$ (Cf. Eq.(8)). Besides, in this particular case
the projected area is temperature-independent, $S = \pi l^2$.

The above considerations bring about the following relationships
between the fitting and the model parametres:
\begin{equation}
 A = \frac{LI_C(0)\alpha }{\mu _0\lambda _L(0)l}, \qquad
 B= \frac{2LI_C(0)\beta}{\mu _0l^2}
\end{equation}
with $\alpha = \Phi _0I_C(0)/2\pi k_BT_C$ and $\beta = U(0)/k_BT_C$.

In conclusion, to check the self-consistency of the model, we recall
that within our scenario, the normalized remanent magnetization
$m_r(t)$ disappears for $T\ge T_J$ (see Fig.1) where $T_J$ is the
phase-locking temperature of the Josephson network. At the same time,
this temperature is usually defined via the equation $U(T_J) =
k_BT_J$ which results in the following two expressions (both valid
near $T_C$) relating the two critical temperatures ($T_C$ and $T_J$):
\begin{equation}
t_J\equiv \frac{T_J}{T_C} = \frac{\alpha}{1+\alpha}
\end{equation}
and
\begin{equation}
(1-t_J)^{3/2} = \frac{t_J}{\beta}
\end{equation}
for YBCO and Nb, and for LSCO based arrays, respectively. Using the
above-mentioned values of the critical temperatures for the three
samples and the experimentally found estimates for $\alpha$ and
$\beta$ (see above), Eqs.(11) and (12) bring about the following
reasonable estimates of the phase-locking temperatures (Cf. Eqs.(4),
(5) and (7)) for the three arrays: $T_J^{YBCO} = (7/8)T_C^{YBCO}$ ,
$T_J^{LSCO} = (1/2)T_C^{LSCO}$ and $T_J^{Nb} = (12/13)T_C^{Nb}$.

In summary, by employing the so-called phase-slip mechanism in each
element of a plaquette with four effective Josephson junctions, we
have consistently derived expressions which lead to a reasonable
description of the temperature behavior of magnetic remanence of
disordered Josephson arrays. The values obtained for the two
experimentally accessible parameters, the locking temperature, $T_J$,
and the superconducting critical temperature, $T_C$, are in good
agreement with the experimental data. The employed approach indicates
that the temperature dependence of the magnetic remanence is
universal, regardless of the origin of the superconducting electrodes
of the junctions which form the array.

\acknowledgements

Brazilian agencies FAPESP, CAPES, PRONEX and CNPq are acknowledged
for partial financial support.

\begin{figure}[htb]
\epsfxsize=8.5cm \centerline{\hspace{0mm} \epsffile{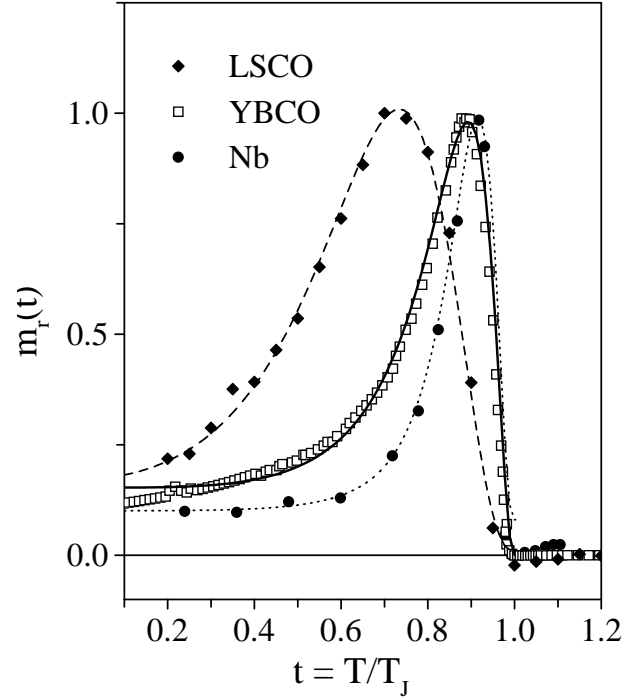}}
\vspace{3mm} \caption{Temperature dependence of the normalized
remanent magnetization $m_r(T)$, showing the experimental data for
three different samples and the corresponding fittings (see text).}
\end{figure}

\end{document}